\documentclass{elsart}

\usepackage{natbib}

\usepackage{amssymb}
\usepackage {graphicx}
\usepackage{placeins}
\usepackage{longtable,array,tabularx}

\begin{document}

\begin{frontmatter}

\title{The host galaxies of low luminosity quasars at high redshift}

\author[tuorla]{J.K. Kotilainen}
\author[padova]{R. Falomo}
\author[como]{A. Treves}
\author[milan]{M. Uslenghi}
\address[tuorla]{Tuorla Observatory, University of Turku, V\"ais\"al\"antie 20, FI--21500 Piikki\"o, Finland (e-mail: jarkot@utu.fi)}
\address[padova]{INAF -- Padova Observatory, Italy}
\address[como]{Universit\`a dell'Insubria, Como, Italy}
\address[milan]{INAF -- IASF, Milan, Italy}

\begin{abstract}
We present VLT/ISAAC near-infrared imaging of the host galaxies of 15 
low luminosity quasars at 1 $<$ z $<$ 2. This work complements our studies to 
trace the cosmological evolution of the host galaxies of 
high luminosity quasars. The radio-loud (RLQ) and radio-quiet (RQQ) quasars 
have similar distribution of redshift and luminosity, and together the high 
and low luminosity quasars cover a large range of the quasar 
luminosity function. Both RLQ and RQQ hosts resemble massive inactive 
ellipticals undergoing passive evolution. However, RLQ hosts are 
systematically more luminous than RQQ hosts, as also found for 
the high luminosity quasars. The difference in the host luminosity remains 
the same from z = 2 to z = 0. For the entire set of quasars, we find 
a correlation between the nuclear and the host luminosities, albeit with 
a large scatter. The correlation is less apparent for the RQQs than for 
the RLQs. 

\end{abstract}


\end{frontmatter}

\section{Introduction}
\label{intro}

Low redshift (z $\leq$ 0.5) quasars are predominantly hosted by massive, 
bulge-dominated galaxies \citep{bahcall97,dunlop03,pagani03}. This is 
consistent with the fact that nearby massive spheroids host inactive 
supermassive black holes (BH) \citep{ferrarese02}, and suggests that 
episodic quasar activity may be common in galaxies and that the nuclear power 
depends on the mass of the galaxy \citep{kauffmann03}.

At low redshift, the BH mass is related to the luminosity and 
velocity dispersion of the bulge \citep{marconi03,bettoni03,haring04}. 
Furthermore, the strong cosmological evolution of quasars is similar to 
the BH mass accretion rate and the evolution of the cosmic 
star formation history \citep{madau98,barger01,yu02}. 
Therefore, determining the properties of quasar hosts close to the peak of 
quasar activity is crucial to investigate the fundamental link between 
the formation and evolution of massive galaxies and the nuclear activity.

The detection of the host galaxies of high redshift quasars is very 
challenging because the host galaxy rapidly becomes very faint compared to 
the nucleus. To cope with this, high spatial resolution and S/N, 
and a well defined PSF are needed. All these requirements can be fulfilled by 
ground-based 8m class telescopes.
We recently carried out a systematic VLT/ISAAC imaging study of 17 quasars 
(10 radio-loud quasars [RLQ] and seven radio-quiet quasars [RQQ]) 
at 1 $<$ z $<$ 2 to characterize their host galaxies \citep{falomo04}. 
The evolution of both RLQ and RQQ hosts until z $\sim$2 is consistent with 
that of massive ellipticals undergoing passive evolution. There is no 
significant decrease in the host mass as would be expected in hierarchical 
formation models \citep{kauffmann00}. RLQ hosts are more luminous by 
$\sim$0.6 mag than RQQ hosts at all redshifts. 

The quasars in \citep{falomo04} belong to the bright end of quasar 
luminosity function. Here we present imaging of quasars that have on average 
lower luminosity by $\sim$2 mag, to study the dependence of host properties 
on nuclear luminosity. We use H$_0$ = 70 km s$^{-1}$ Mpc$^{-1}$, 
$\Omega_m$ = 0.3 and $\Omega_\Lambda$ = 0.7.

\section{Sample, observations and analysis}
\label{obs}

The low luminosity quasars were extracted from \citet{veron03}, 
requiring 1.2 $<$ z $<$ 1.9, $M_B <$ -25.5 at z $<$ 1.5 up to 
$M_B <$ -26.7 at z = 1.9, and having bright stars within 1 arcmin of 
the quasar to allow a reliable characterization of the PSF. Nine RLQs and 
six RQQs were observed. 
Their redshift and luminosity distributions are shown in Fig. 1. 
They are matched in redshift with the high luminosity quasars 
\citep{falomo04}, and together the samples cover a large fraction 
($\sim$4 mag) of the quasar luminosity function. 

\begin{figure}[h!]
\begin{center}
\includegraphics[width=10cm]{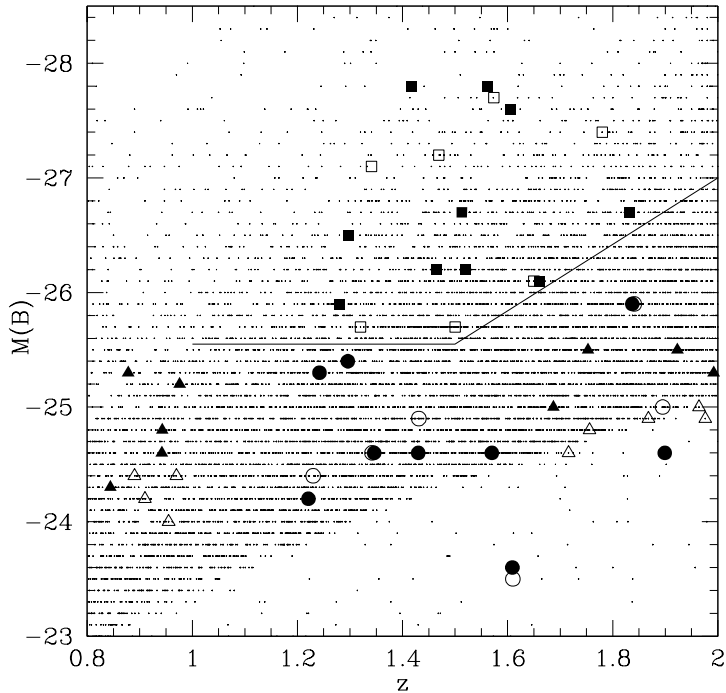}
\caption{
Low luminosity RLQs (filled circles) and RQQs (open circles), high luminosity 
RLQs (filled squares) and RQQs (open squares) from Falomo et al. (2004), and
RLQs (filled triangles) and RQQs (open triangles) from HST/NICMOS study by 
Kukula et al. (2001) in the z - M$_B$ plane, compared with all quasars 
(small dots) in Veron-Cetty \& Veron (2003). The solid line shows 
the luminosity limit applied between the low and high luminosity samples. 
}
\label{fig:MBz}
\end{center}
\end{figure}

The quasars were imaged in the $H$- or $K$-band using VLT/ISAAC, 
with total integration time of 36 minutes per target. The seeing was excellent 
during the observations (median $\sim$0.4 arcsec FWHM).
Images were taken using random jitter with 2 minute exposure per frame, 
Pipeline data reduction included flat fielding, median sky subtraction and 
co-addition of aligned frames. 

For each field, we performed a 2D analysis of all stars to construct 
a composite PSF model. For each quasar, we measured its 
2D luminosity distribution, masking regions affected by companions. 
The luminosity distributions were modeled, using an iterative 
least-squares fit, into a point source (PSF model) and an elliptical galaxy 
(r$^{1/4}$), convolved with the PSF. 

\section{Results}
\label{results}

\subsection{The evolution of quasar hosts}
\label{evol}

For all the low luminosity quasars, except one RQQ, the host galaxy is 
resolved. The average absolute magnitude of their host galaxies is 
$<M_K>$(host) = {--26.3 $\pm$ 0.3} (RLQ) and 
$<M_K>$(host) = {--25.7 $\pm$ 0.3} (RQQ), compared with those for 
high luminosity quasars \citep{falomo04}: 
$<M_K>$(host) = {--26.8 $\pm$ 0.2} (RLQ) and 
$<M_K>$(host) = {--26.0 $\pm$ 0.3} (RQQ). 

Fig. 2 shows the average luminosities of the host galaxies in quasar samples 
at z $<$ 2, from HST or ground-based data. The host galaxies of both RLQs and 
RQQs, despite their different radio properties, follow the passive evolution 
of massive ellipticals. Similar luminosity evolution is seen in radio galaxies 
at least out to z $\sim$2.5 \citep{pentericci01,willott03}. 
This passive evolution is also consistent with spectroscopic studies of 
low redshift quasar hosts and radio galaxies \citep{devries00,nolan01},  
indicating that their stellar content is dominated by an old 
stellar population. 

\begin{figure}[h!]
\begin{center}
\includegraphics[width=10cm]{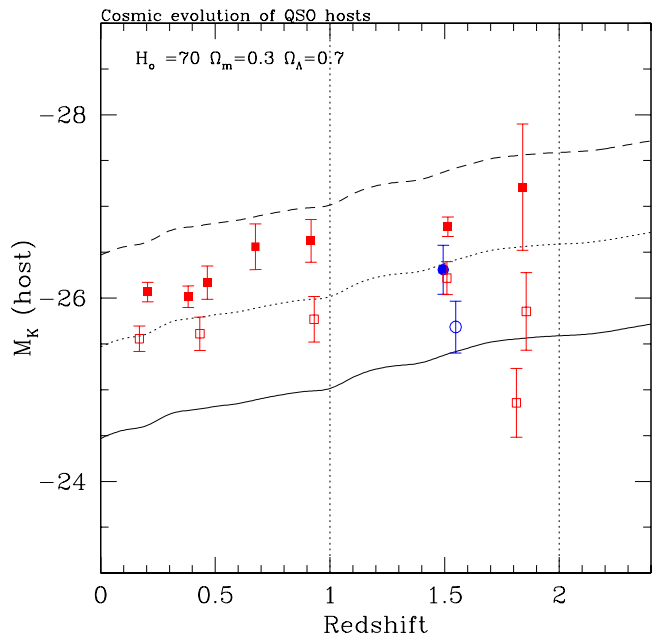}
\caption{
The evolution of quasar host luminosity. The lines show the expected behavior 
of a massive elliptical (at M$^*$, M$^*$-1 and M$^*$-2; 
solid,dotted and dashed line) undergoing passive stellar evolution 
\cite{bressan98}. 
}
\label{fig:evol}
\end{center}
\end{figure}

The cosmic evolution traced by quasar hosts up to z $\sim$2 disagrees with 
semianalytic hierarchical models of AGN and galaxy formation and evolution 
\citep{kauffmann00}), which predict fainter (less massive) hosts at 
high redshift, which merge and grow to form low redshift massive spheroids. 
Thus, if quasar hosts undergo passive evolution, it is likely that their 
mass remains essentially unchanged from z $\sim$2 up to z = 0.

For both low and high luminosity quasars, RLQ hosts are systematically 
more luminous than RQQ hosts by $\sim$0.6-0.8 mag. There is no significant
change in this luminosity gap with redshift.
If the host luminosity is correlated with the BH mass, this indicates 
that the BHs in RLQs are more massive than those in RQQs. 
Furthermore, to produce the same luminosity, the BHs in RLQs must be accreting 
less efficiently than those in RQQs.

\subsection{Nuclear versus host properties}
\label{nuchost}

In Fig. 3 we compare the $K$-band host and nuclear luminosities of the low 
and high luminosity quasars. For the entire set of quasars, we find 
a reasonably strong nuclear luminosity dependence of the host galaxy 
luminosity, with a rather large scatter. Separating the two quasar types, 
a modest correlation can still be seen for RLQs, which is hardly present 
for RQQs. Since no or little correlation has generally been found at 
low redshift \citep {dunlop03,pagani03}, it must have its onset at relatively 
high redshift. 

\begin{figure}[h!]
\begin{center}
\includegraphics[width=10cm]{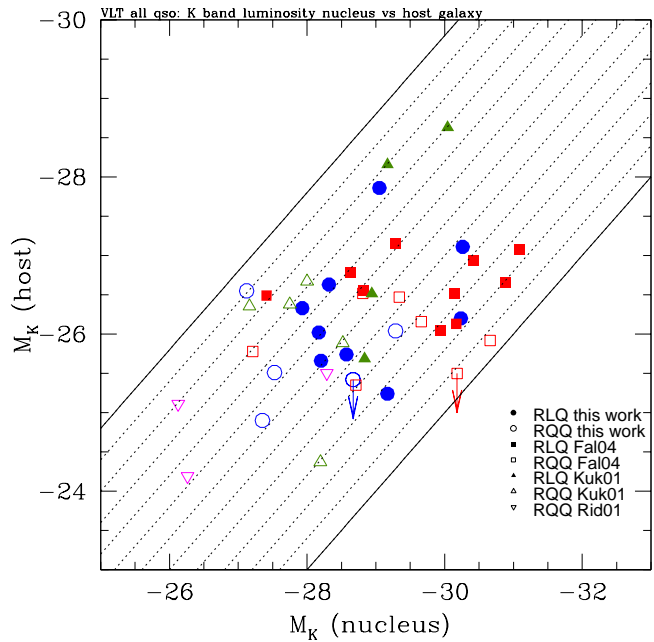}
\caption{
The absolute magnitude of the nucleus compared with that of the host galaxy. 
For symbols, see Fig. 1. Diagonal lines represent the loci of constant 
N/H ratio. Separations between dotted lines correspond to a difference of 
factor of 2 in N/H, while the solid lines encompass a spread of 1.5dex in N/H.
}
\label{fig:nuchost}
\end{center}
\end{figure}

Assuming that the $K$-band host luminosity is proportional to the BH mass 
(as observed at low redshift) and that the $K$-band nuclear luminosity is 
proportional to the bolometric luminosity, the $K$-band nucleus/host (N/H) 
luminosity ratio is proportional to the Eddington ratio L/L$_E$. 
For low luminosity quasars, the average log N/H ratios are: 
1.04 $\pm$ 0.40 (RLQ) and 1.00 $\pm$ 0.44 (RQQ), compared with those for 
high luminosity quasars \citep{falomo04}: 1.36 $\pm$ 0.33 (RLQ) and 
1.32 $\pm$ 0.37 (RQQ). 
This indicates that high redshift quasars radiate with a wide range 
of L/L$_E$, irrespective of radio power. Similarly wide range was found for 
low redshift RLQs \citep{pagani03}, suggesting that the spread does not 
depend on redshift either. This is consistent with the evolution of quasars 
being mainly produced by a density evolution of BH activity due to increased 
merger and fuelling rate at high redshift. L/L$_E$ appears, however, to depend 
on quasar luminosity. This indicates a smaller accretion efficiency in 
low luminosity quasars. 

\section{Open questions}
\label{open}

To determine quasar host properties at even higher redshift, close to the peak 
epoch of quasar activity (z $\sim$2.5) and beyond, requires very 
high S/N observations and a very narrow reliable PSF. We have an ongoing 
program \citep{falomo05} to tackle this using adaptive optics imaging 
with VLT/NACO (see also Falomo et al., in these proceedings). 
Colour information for the hosts, spectroscopy to estimate BH masses, 
and environments as a function of redshift and radio power will also be 
topics of future work. 




\end{document}